\documentclass[epsfig,preprint2]{aastex}

\shorttitle{``Soft X-ray transients'' which aren't soft}
\shortauthors{Brocksopp et al.}

\begin{document}

\title {``Soft X-ray transient'' outbursts which aren't soft}
\author{C. Brocksopp}
\affil{Astrophysics Research Institute, Liverpool John Moores University, Twelve Quays House, Egerton Wharf, Birkenhead CH41 1LD, UK}
\affil{Mullard Space Science Laboratory, University College London, Holmbury St. Mary, Dorking, Surrey~RH5~6NT,~UK}
\email{cb4@mssl.ucl.ac.uk}
\author{R.M. Bandyopadhyay}
\affil{Dept. of Astrophysics, University of Oxford, Keble Road, Oxford OX1 3RH, UK}
\author{R.P. Fender}
\affil{Astronomical Institute ``Anton Pannekoek'' and Center for High-Energy Astrophysics, University of Amsterdam, Kruislaan 403, 1098 SJ Amsterdam, The Netherlands\\}

\begin{abstract}
We have accumulated multiwavelength (X-ray, optical, radio) lightcurves for the eight black hole X-ray binaries which have been observed to enter a supposed `soft X-ray transient' outburst, but remained in the low/hard state throughout the outburst. Comparison of the lightcurve morphologies, spectral behaviour, properties of the quasi-periodic oscillations and the radio jet provides the first study of such objects as a sub-class of X-ray transients. However rather than assuming that these hard state X-ray transients are different from the `canonical' soft X-ray transient, we prefer to consider the possibility that new analysis of both soft and hard state X-ray transients in a spectral context will provide a model capable of explaining the outburst mechanisms of (almost) all black hole X-ray binaries.
\end{abstract}


\section{Introduction}

With the list of known soft X-ray transients (SXT) growing longer every year and now totalling $\ge30$ it is inevitable that we should find a few sources which do not follow the classical behaviour that we expect. In particular, there are a number of supposed black hole SXTs which do not show a soft component in their X-ray spectra. Instead they display power-law emission across the X-ray energy spectrum, despite some sources reaching high luminosities.

Models for SXT outbursts predict the presence of a thin accretion disc to explain the soft black-body spectrum emitted by the lower energy X-rays (e.g. Esin et al. 1997). The power-law emission at higher energies is thought to be produced via reprocessing in some form of hot plasma (e.g. corona or advection dominated accretion flow), which surrounds the compact object and upscatters disc and/or stellar photons to higher energies via the inverse Compton emission mechanism. This soft spectrum has indeed been observed in the majority of soft X-ray transient outbursts.

\subsection{The low/hard state}

Some sources, however, show power-law emission down to low X-ray energies and this has been the case for the sources considered in this paper; despite entering outburst there has been no evidence for a soft blackbody component at X-ray wavelengths. In these objects it is assumed that all X-ray emission is produced via reprocessing in the `corona' and not in the accretion disc.

The power-law X-ray energy spectrum is typical of black hole candidates in the low/hard state, most notably Cyg X-1 which spends most of the time in this state (Brocksopp et al. 1999). Corresponding X-ray power spectra for these sources show a high level (relative to the other spectral states) of low-frequency noise, a broken power-law and at least one quasi-periodic oscillation. (See e.g. van der Klis (1995) for further definition of spectral states.) The state is also characterised by the presence of a low-intensity, but extremely powerful, radio jet which emits synchrotron radiation at radio and often higher frequencies (Fender 2001; Corbel \& Fender 2002). During the high/soft state we see the presence of the soft black-body component in the X-ray spectrum and the quenching of the radio jets, although rapid radio-emitting ejections appear to accompany changes in X-ray spectral hardness (e.g. Brocksopp et al. 2002) -- it is this behaviour that we expect in the classic black hole SXT events. 

In this paper we take a closer look at eight black hole `SXTs' which, on the basis of their observed X-ray and radio properties, have been reported to have spent the whole of their outburst in the low/hard state. Although observing coverage was incomplete for the less recent outbursts, each source (other than 4U 1543$-$475) was observed at low X-ray energies for the whole outburst or at least a month after the initial rise, by which time the decay had started and no soft disc component had been discovered; while it is common for SXTs to spend some time in the low/hard state prior to softening, this period is generally only a few days (the longest initial low/hard state period observed to date is that of XTE J1650$-$500 which remained hard for $\sim 1$ month before softening but also continued to rise in flux; Wijnands priv. comm.). Thus it is unlikely that any of the sources made an unobserved transition to a softer state. The sources are listed in Table~\ref{sources} along with the date of the low/hard state outburst, the observations which took place and the sources from which we obtained data-points; the observing coverage at the various different frequencies can be seen in the lightcurves plotted in Fig.~\ref{lightcurve1}. A brief summary of the properties of each source, in chronological order of low/hard state outburst, is given below. 

\subsection{V404 Cyg}

V404 Cyg is probably one of the best-known X-ray transient sources on account of its high luminosity and levels of variability (at various wavelengths) during both outburst and quiescence (e.g. Tanaka \& Lewin 1995, Hynes et al. 2002) and many observations exist (e.g. Han \& Hjellming 1992, Shahbaz et al. 1994). Despite the high luminosity of the outburst, the lack of soft component present in the X-ray spectrum suggests that it was indeed a low/hard state transient (e.g. Oosterbroek et al. 1997) -- we note that this has been disputed and that it may have spent a very short period in the soft state (\.Zycki et al. 1999) but include it here for completeness. 

\subsection{A1524$-$62}
A1524$-$62 has been observed in outburst twice$^{1}$, the first of which was a `normal' SXT event (Kaluzienski et al. 1975). The second outburst was observed in the hard X-ray regime by SIGMA (Barret et al. 1992) and by chance was also observed at soft X-rays during the {\it ROSAT} All Sky Survey ten days before the SIGMA observation. The soft X-ray spectrum could be equally well fit by either a cool black body or a power-law; the presence of an ultra-soft component could not be confirmed or ruled out (Barret et al. 1995). We note, however, that the outburst was insufficient to trigger the soft X-ray all sky monitors on-board {\it WATCH} and {\it Ginga}, the latter providing an upper limit of 50 mCrab which was consistent with the ROSAT detection (Barret et al. 1995). Thus we include this source as a possible hard state candidate.

\subsection{4U 1543$-$475}
4U 1543$-$475 is another recurrent transient which has been observed in outburst four times\footnote[1]{Due to the poor positional accuracy of the initial observations of A1524$-$62, 4U 1543$-$475 and GS 1354$-$64, it is of course possible that the earlier outbursts of these systems were in fact additional sources, as discussed by e.g. Brocksopp et al. (2001) and Kitamoto et al. (1987). However, given that e.g. XTE J1550$-$564, GX 339$-$4 have indeed been observed recently to exist in a number of spectral states (Corbel et al. 2001, 2003) there is currently no reason to assume that these were necessarily different sources, particularly as they lie in relatively uncrowded regions.} and was a classical SXT during the first two of these, as well as the fourth and current outburst (Matilsky et al. 1972, Kitamoto et al. 1984, Miller \& Remillard 2002). During the third event hard X-ray observations revealed a power-law spectrum (Harmon et al. 1992) but since there were no soft X-ray observations we cannot confirm the presence or otherwise of a soft component at lower frequencies. We include this source as a possible hard state candidate.

\subsection{GRO J0422+32}
GRO J0422+32 was first observed in 1992 when the X-ray source entered a `canonical' fast rise, exponential decay type outburst (Paciesas et al. 1992). The energy spectrum was a hard power-law (Sunyaev et al. 1994) and timing analysis revealed properties consistent with the low/hard state, although the expected anti-correlation between break frequency and power at the break was not present (van der Hooft et al. 1999). The radio counterpart and the evolution of its spectrum throughout the event are discussed in Shrader et al. (1994). The optical counterpart was discovered by Castro-Tirado et al. (1993) and the intense optical monitoring which followed resulted in the detection of two additional and unusual `mini-outbursts' (e.g. Callanan et al. 1995).

\subsection{GRO J1719$-$24}
GRO J1719$-$24 was detected simultaneously by BATSE and SIGMA and was notable for the stability of the emission at hard X-ray wavelengths (Harmon et al. 1993, Ballet et al. 1993). The low/hard state nature was confirmed through study of the X-ray energy spectra and found to be comparable with that of Cyg X-1 in the low/hard state (Revnivtsev et al. 1998); the power spectra showed the presence of a QPO which varied in frequency (van der Hooft et al. 1996). The radio and optical counterparts were discovered by Della Valle et al. (1994).

\subsection{GRS 1737$-$31}
GRS 1737$-$31 was discovered by SIGMA (Trudolyubov et al. 1999) and was also observed by $RXTE$/PCA, which confirmed the presence of the hard power-law spectrum at lower energies (Cui et al. 1997). Further X-ray observations were made by {\it BeppoSAX} and {\it ASCA} (Heise 1997, Ueda et al. 1997) which improved the positional accuracy and confirmed that the hard spectrum was still present (at $\sim$ 2--20 keV) $\sim$ 2 weeks after the initial detection; there were no observations at optical or radio wavelengths.

\subsection{GS 1354$-$64}
GS 1354$-$64 is another recurrent transient and it has been observed during four different outbursts$^{1}$ (Brocksopp et al. 2001). Assuming that all events were indeed from the same source, this was first ever SXT outburst observed (then named Cen X-2) and was also one of brightest (e.g. Tanaka \& Lewin 1995, and references therein). The first and third events were characterised by very soft X-ray spectra; the second and fourth were hard events (Brocksopp et al. 2001 and references therein) -- on account of the observing coverage we consider just the most recent event in this paper, during which a hard power-law energy spectrum was observed throughout by {\sl RXTE} (Revnivtsev et al. 2000).

\subsection{XTE J1118+480}
XTE J1118+480 was discovered in 2000 during an outburst (Remillard et al. 2000); the presence of the low/hard state was determined from study of the hard power-law X-ray energy spectrum and the characteristic power-law, high rms noise and QPO in the X-ray power spectrum (Revnivtsev et al. 2000a). The optical and radio counterparts were also discovered during the outburst (Uemura et al. 2000, Pooley 2001, Fender et al. 2001).

\section{Lightcurves}

Lightcurves have been constructed for each of the eight sources that has exhibited a low/hard state outburst. The data points have been obtained from the literature and the {\sl RXTE}/ASM and {\sl CGRO}/BATSE web-sites (see Table~\ref{sources}). We note that in a number of cases, datasets have been measured using the ADS software {\sc dexter}; since there is a certain amount of error involved in making these measurements we recommend that the points shown here are used as a guide and the original papers consulted. For the same reason (and also due to lack of availability) a number of our plots do not show error bars -- again these can be seen in some of the original papers.

Although all data displayed here have been published previously, it has been uncommon for all wavebands to be analysed in conjunction with each other (for combined X-ray/optical/radio analyses see e.g. Brocksopp et al. 2001, Han \& Hjellming 1992, McClintock et al. 2001). Therefore for each source we have attempted to find data at X-ray (in more than one energy band), optical and radio wavelengths, although this was not always possible. Fig.~\ref{lightcurve1} shows X-ray lightcurves for A1524$-$62, 4U 1524$-$475 and GRS 1737$-$31 and X-ray, optical and radio lightcurves for V404 Cyg, GRO J1719$-$24 and GRO J0422+32. Fig.~\ref{lightcurve2} (contd.) shows X-ray, optical and radio lightcurves for GS 1354$-$64 and XTE J1118+480. The X-ray intensity values have all been converted into Crab units according to the approximate conversions presented in Table~\ref{conversions}. Optical and radio points are shown in milli-Janskys.

It is immediately apparent that each source shows very different types of behaviour. The fast-rise exponential-decay (FRED) lightcurve that is considered the `classic' signature of a soft X-ray transient (see e.g. Chen et al. 1997) is not predominant in a number of these lightcurves; only in V404 Cyg is the lightcurve morphology of both X-ray and optical emission `canonical', although this being true also for A1524$-$62, 4U 1543$-$475 and GRS 1737$-$31 cannot be ruled out due to lack of data. The X-ray, but not the optical, lightcurve of GRO J0422+32, shows a FRED, as does the `precursor' X-ray lightcurve of XTE J1118+480; the second and more significant outburst of XTE J1118+480 (X-ray and optical) shows a `plateau' lightcurve morphology, as did the initial maximum of GRO J1719$-$24. Continuing to use the nomenclature of Chen et al. (1997), the X-ray lightcurves of GS 1354$-$64, the subsequent maxima of GRO J1719$-$24 and possibly A1524$-$62, 4U 1543$-$475 and GRS 1737$-$31 show `triangular' morphologies. The optical lightcurves of GRO J0422+32 and GS 1354$-$64 are indeterminate although possibly some variation on a FRED shape. The radio lightcurves, although sometimes lacking in observing coverage, appear to display typical jet ejection lightcurves as discussed in Fender (2001) and the references listed in Table~\ref{sources}.

Each X-ray outburst, except GS 1354$-$64 (and A1524$-$62?), is associated with at least one additional secondary outburst either during the decay (i.e. V404 Cyg, GRO J0422+32) or following return to quiescence by a few tens to a few hundreds of days (GRO J1719$-$24, XTE J1118+480, GRS 1737$-$31) -- some of these secondary maxima have been classified as a `glitch', `bump' or `mini-outburst' by Chen et al. (1997). The main outburst of GRO J1719$-$24 was followed by 4--5 smaller peaks, reminiscent of the 1994 outburst of GRO J1655$-$40 (e.g. Harmon et al. 1995), although the latter was much brighter and showed more dramatic variability.

The optical lightcurves are not as well-correlated with the X-rays as we might expect considering that the optical light is thought to be due to the reprocessing of X-rays, although the general trends are often similar. It is not clear whether or not the secondary maximum seen in the V404 Cyg optical lightcurve occurs also in the X-ray and the few points available suggest that the optical maximum of GRO J1719$-$24 was short-lived, unlike the X-ray although they appeared to rise (quasi-) simultaneously. The optical peak of GRO J0422+32 may well have taken place prior to the X-ray peak and then decayed much more slowly than the X-rays; following the X-ray return to quiescence, the optical flux density dropped more rapidly before increasing $\sim 100$ days later in the first of two additional peaks, unaccompanied by {\em any} X-ray behaviour. The partial optical lightcurve of GS 1354$-$64 showed a decay while the X-rays were still rising, a peak plus its decay and then a further increase (this is discussed in detail in Brocksopp et al. 2001). Conversely, the optical lightcurve of XTE J1118+480 does seem to be reasonably well-correlated with the X-rays.

Finally, where coverage is available, the main outburst and the subsequent secondary maxima of the X-ray lightcurves tend to be associated with radio ejections, although this is {\em not} to say that the lightcurves correlate, even when taking time-lags into account; they do not correlate and this would not necessarily be expected (Kuulkers et al. 1999). In the case of XTE J1118+480 the maximum radio intensity was maintained during the extended X-ray peak, reminiscent of the low/hard state jet of Cyg X-1 and GX 339$-$4 (e.g. Fender 2001). 


The most notable feature of these lightcurves is their inconsistency -- particularly considering the similarities of their X-ray (and broad-band, see Section 5) spectral behaviour. If we had considered just the multiwavelength lightcurves then it is unlikely that these sources would have been labelled a `class of object' distinct from the genuinely soft X-ray transients. These plots reveal a number of questions yet to be answered by current models. Why are some of the outbursts FRED-type at one wavelength but not at others? Why are such different lightcurves produced in such (X-ray) spectrally similar sources? Why do the secondary outbursts take place on such variable timescales? Why are some of the outbursts (un)correlated with other wavebands? Why are the durations of the peaks so variable? 

Most of these questions remain unanswered because, in applying the disc instability model (or other models) to X-ray transients, we have failed to consider the full multiwavelength aspect. The dynamical effect of jets (such as removal of power and angular momentum -- the jets can potentially account for up to 90\% of the accretion power; Fender 2001) on the models has often not been accounted for; this is particularly important since the jet is thought to emit in the infrared and perhaps optical (and X-ray?) regimes as well as the radio (see Section 5). It has further been assumed that `canonical' X-ray transients follow the FRED morphology whereas this is frequently not the case, particularly in the X-ray lightcurves; even in the genuinely soft sources a precursor low/hard state phase is seen prior to the fast rise in all cases where data is available (Brocksopp et al. 2002). We investigate alternatives to these previous asssumptions in Section 6.

\section{Spectral Evolution}

It might be expected that X-ray transients would soften as they increase their flux during an outburst and then harden again during the decay -- similar to the spectrum of Cyg X-1 during its well-known state-change of 1996 (e.g. Zhang et al. 1997). Therefore it was somewhat surprising to see that the SXT XTE J1859+226 was very hard at the beginning of the outburst (peaking first in the hard X-rays) and gradually softened throughout the decay (Brocksopp et al. 2002). Superimposed upon this trend were small-scale temporary hardenings which coincided with radio jet ejections. Likewise XTE J1550$-$564 also showed a general softening as the luminosity decayed following the 1998 outburst (Wu et al. 2002), although interestingly the 2000 outburst hardened during decay (Soria et al. 2001). Thus a correlation between luminosity and spectral softness is not necessarily observed in all cases.

While ideally we would have studied the variability of the XTE/ASM hardness ratio data for the low/hard state sources, this has proved unreliable due to the poor signal-to-noise ratio; XTE J1118+480 showed a possible softening at the onset of the outburst but this was non-conclusive -- however GRO J1719$-$24 did indeed soften as the flux rose above quiescence (Revnivtsev et al. 1998). Revnivtsev et al. (2000b) found a general softening in the spectra of GS 1354$-$64 over the course of the outburst, with a sudden hardening in the final observation (during the decay but prior to quiescence) which coincided with a possible increase in flux. The TTM hardness ratio of GRO J0422+32 remained approximately constant (Sunyaev et al. 1994) but we note that none of these observations took place during the rise to outburst (which is where any significant spectral changes would be expected to take place). The spectral behaviour of V404 Cyg is more complicated than the other sources and further discussion is beyond the scope of this paper -- instead we refer the reader to Oosterbroek et al. (1997) and \.Zycki et al. (1999).

\section{The low-frequency QPO}

The presence of a quasi-periodic oscillation (QPO) in the X-ray power spectrum of an X-ray binary is a common feature of the low/hard state, as well as the intermediate and very high states (e.g. Wijnands \& van der Klis 1999 and references therein) -- however, the frequencies at which they occur seem to be noticeably lower in the low/hard state. This has been the case in V404 Cyg, GRO J1719$-$24, GRO J0422+32, GS 1354$-$64 and XTE J1118+480, all of which show a QPO at  mHz frequencies (see Oosterbroek et al. 1997, van der Hooft 1996, 1999, Brocksopp et al. 2001 and Wood et al. 2000 respectively) -- furthermore in each case  other than V404 Cyg (for which a QPO at $\sim 0.05$ Hz was detected during the brightest observation only) the frequency of the QPO does not remain constant. Instead it increases during the outburst, showing little or no apparent correlation with the X-ray intensity; the presumed decrease in frequency following the return to quiescence has not been observed to date for any of the low/hard state sources.

Plots showing the frequency variability of the QPO in each source are shown in Fig.~2. We note that in the case of GRO J0422+32 there were two different QPOs detected, at approximately 40 and 200 mHz; it is the latter which shows the variability. It is also the only source for which there is even the hint of a correlation between the QPO frequency and the X-ray flux -- however, cross-correlating the QPO data with the X-ray lightcurve does not confirm the presence of any correlation in {\em any} of the sources. Furthermore we find that the duration of the QPO frequency increase is a factor of at least 1.5--2 times the duration of the X-ray flux rise (with the possible exception of XTE J1118+480 for which the rise times may be comparable, although there is insufficient data to be certain). The QPO frequencies of GS 1354$-$64 are notably lower (by a factor of $\sim 10$) than those of the other sources; we note that the outburst of GS 1354$-$64 was noticeably less dramatic that the others, although it is not at all clear whether this is related to the lower frequency of the QPO. It is also possible that, as in the case of GRO J0422+32, a second QPO was present (but undetected) in GRO J1719$-$24, GS 1354$-$64 and XTE J1118+480 which would explain the difference between the frequencies.

The remaining three sources have not been reported to display QPOs. In the case of A1524$-$62 and 4U 1543$-$475 this is because there is no published power spectrum. In the case of GRS 1737$-$31 a power spectrum has been computed but the data quality insufficient to determine whether or not a QPO was present (Cui et al. 1997, Cui 2001 priv. comm.).

It has been suggested that the mHz QPO of X-ray transients may relate to the inner edge of the accretion disc (e.g. Revnivtsev et al. 2000b) -- thus if the softening during the outburst is due to the inner edge of the disc moving inwards we would expect an increase of the QPO frequency, as indeed we see (Revnivtsev et al. 2000b). This scenario seems feasible, especially since during genuinely soft outbursts the sources tend to display higher frequency (Hz-kHz) QPOs (e.g. Cui et al. 2000). However, we would also expect the inner edge of the disc to move outwards again at the return to quiescence, thus decreasing the QPO frequency back to its ``quiescent value''; while this behaviour has not been observed in any of these sources (although hinted at in GRO J0422+32) it has been observed during the low/hard state phases of the 2000 outburst of XTE J1550$-$564 (Reilly et al. 2001).

An alternative origin for this QPO is presented in Wood et al. (2001). They suggest that it is produced at a large radius within the disc and that its frequency is inversely proportional to the disc mass. While a decrease in QPO frequency after the outburst is still expected, their model predicts this to take place {\em after} the return to quiescence; measurements of the QPO frequency have not yet taken place for long enough after the peak (for any source) for this to have been observed. However, if this model is correct then we might also expect a similar relationship between the disc mass and QPO frequency in canonical soft X-ray transients. In the case of these objects the QPO frequency is higher and so an even greater reduction in disc mass during the outburst would be required; this would depend on the relative amounts of mass transfer and mass accretion taking place as the inner edge of the accretion disc moves inwards to the last stable orbit (as is predicted to take place during the transition to softer states e.g. Esin et al. 1997).

\section{`Radio jets' and optical emitting regions}

Evidence for the ubiquity of radio jets in most classes of X-ray binaries is mounting constantly. As well as probably all black hole candidates in the low/hard state and during outburst, the Z-sources (whilst on the horizontal branch) and the transient neutron star systems (whilst in outburst) also show evidence for some form of jet behaviour (e.g. Fender 2001, Fender \& Hendry 2000). However, it is also apparent that the jet is not purely a `radio jet' -- the synchrotron spectrum that is thought to be the jet signature is frequently seen up to higher frequencies in the infrared and possibly the optical. In the case of XTE J1118+480 Markoff et al. (2001) were able to fit a synchrotron spectrum to the full range of frequencies up to the hard X-rays (with an additional black-body disc component in the optical).  More recently a similar model has been successfully applied to thirteen broad-band spectra of differing luminosities observed from GX 339$-$4 during the low/hard state. The jet model also analytically predicts the slope of the radio/X-ray correlations seen in this source (Markoff et al. 2003, Corbel et al. 2003). If the power-law X-ray emission is indeed emitted by the jet then these jets are considerably more powerful and use a much greater fraction of the accretion energy than most current models allow.

A flat or inverted synchrotron spectrum has already been well-established at radio frequencies in each of the low/hard state sources studied here (where radio observations were available, Fender 2001). We have therefore compiled the full spectrum from radio to X-rays in order to determine whether the synchrotron spectrum of XTE J1118+480 is an exception or the rule. The resulting broad-band spectra for V404 Cyg, GRO J1719$-$24, GRO J0422+32 and GS 1354$-$64 are shown in Fig.~\ref{spectrum} for three different epochs per source. The epochs were chosen so as to make optimum use of simultaneous observations and also to sample a range of different types of activity -- for example, the peak of the initial outburst, a secondary outburst and a period of low luminosity (this was not possible for GS 1354$-$64 for which the three epochs were the rise, peak and decay). All optical points have been de-reddened according to the extinction values quoted in Liu et al. (2001) and the interstellar extinction curve of Whittet (1992).

Initial inspection of the resultant broad-band spectra suggests that despite the very different lightcurve properties of each outburst, their spectral properties were similar. It is also notable that the spectra at different epochs do not vary considerably. The flat/inverted radio behaviour is clearly seen for each source, although some epochs are contaminated by optically thin ejections (e.g. GRO J1719$-$24: JD 2449777), but as is so often the case (Fender 2001) it is not possible to determine the high energy cut-off to the synchrotron spectrum. While the frequency coverage is not good enough to confirm that the synchrotron spectrum extends to the X-ray regions, neither can we state categorically that it is impossible; the spectra of these four sources here resemble that of XTE J1118+480 and formal fitting of these broad-band spectra with a synchrotron spectrum is in progress\footnote[1]{Preliminary results suggest that some of these spectra can indeed be fit with the Markoff et al. (2001) jet model.}.

It is important to note that while much of this work is speculative on account of insufficient observations in the millimetre regions of the spectrum, it is not always possible to fit a disc spectrum to the optical emission. The most notable example is XTE J1118+480 which requires an additional flat-spectrum component, possibly synchrotron (Hynes et al. 2000). Likewise GRO J0422+32 was poorly fit by a disc spectrum and was {\em better} fit by a self-absorbed synchrotron spectrum (Hynes \& Haswell 1999). Although in this latter case the fit was determined when the source was close to quiescence, as we noted above this need not have much effect since the spectrum was approximately constant throughout the outburst period (Fig.~\ref{spectrum}). We acknowledge that an ADAF is also expected to emit optical and UV synchrotron emission (e.g. Esin et al. 1997) but it is remarkable that the flux density of both jet and ADAF should be apparently so similar. There must be a very close relationship between the two components, if indeed they co-exist (e.g. Meier 2001).

In support of this we note that in the case of GS 1354$-$64, XTE J1118+480 and V404 Cyg the optical:soft X-ray luminosity ratio was surprisingly low ($\le 19$ compared with an average of $\sim 22\pm 1$; van Paradijs \& McClintock 1995) relative to that of soft X-ray transients, thus suggesting that X-ray processing was not the only (or even the dominant) source of optical emission, particularly as some sources peak in the optical {\em before} the X-ray (Brocksopp et al. 1999, Chen et al. 1997). Inspection of the broad-band spectra would suggest that the optical:soft X-ray luminosity ratio was probably also low in GRO J1719$-$24 and GRO J0422+32. While it is possible that the excess optical emission could be due to the companion star and/or accretion stream impact point, again it is a remarkable coincidence that the radio and optical flux densities of each source are so comparable (Fender 2001), particularly in those sources where there is also some degree of correlation between the optical and radio lightcurves.

However, there are further features which may indicate that the broad-band spectrum is not always dominated by the jet, particularly as the optical and radio lightcurves are not necessarily well-correlated. Brocksopp et al. (2001) show that in the case of GS 1354$-$64 the various peaks of the optical lightcurve implied that the increased mass flow (and/or instability) which caused the outburst was observed directly as it crossed the optical-emitting regions of the disc, suggesting that the optical luminosity excess (relative to the X-ray) is not necessarily synchrotron emission. Furthermore the QPO discovered in XTE J1118+480 at optical frequencies was found to be correlated with the X-ray data (Kanbach et al. 2001) but with an optical dip preceding the X-ray peak that is not well-explained by reprocessing of X-rays (Spruit \& Kanbach 2002); instead it was suggested that a relatively slow, dense outflow from the inner regions of the accretion disc was present, producing optical cyclosynchrotron emission. If we are to explain the broad-band spectra in terms of a dominating synchrotron jet then further investigation of the nature of the QPO would be necessary, particularly at optical wavelengths.

\section{Discussion}

The aim of this paper is not necessarily to `drive a wedge' between the `canonical' black hole SXTs and this sub-class of low/hard state X-ray transients (LHXT; we note that the term `hard X-ray transient' has been applied previously to the Be + neutron star binary systems). While we can see that the sources studied here are not classical FRED sources, neither are a large number of SXTs, on account of both lightcurve morphology (Chen et al. 1997) and X-ray spectral state (e.g. Brocksopp et al. 2002); thus it is becoming timely for a re-definition of the `canonical' X-ray transient which incorporates the now numerous radio jet and spectral (X-ray and broad-band) observations. 

Models for SXT outbursts have been largely based on the disc instability model (DIM), initially developed for the outbursts of dwarf novae but later adapted for SXTs (e.g. Lasota 2001 and references therein). The model's success has been due to its ability to produce FRED-shaped lightcurves, thus having the potential to describe observations of X-ray transient outbursts (e.g. King \& Ritter 1998; Dubus et al. 2001). The original model has been adapted to include additional physical mechanisms of relevance in X-ray binaries; by invoking disc irradiation and truncation of the inner disc, the basic rise, decay and recurrence times can be reproduced (Lasota 2001).

There are a number of problems with the DIM, particularly as a large number ($>50\%$; Chen et al. 1997) of SXT outbursts do not display `canonical' FRED behaviour. These problems are well-known (e.g. Chen et al. 1997) and it is thought that they may be rectified with the inclusion of accretion disc coronae and/or warps (Lasota 2001). Additionally there are difficulties in describing the quiescent and low/hard states without the inclusion of an ADAF or coronal outflow (e.g. Esin et al. 1997, Merloni \& Fabian 2002). However it is also of vital importance that the broad-band, and particularly the X-ray spectra are taken into account, especially if a truncated inner disc is required in order to model the lightcurves. In particular the DIM assumes that the X-ray outburst takes place in the disc and that the optical emission is produced via reprocessing of these soft X-ray photons (e.g. Lasota 2001) -- but in the case of the outbursts which remained in the low/hard state all X-ray photons were produced in the corona; the DIM does not consider the production of this power-law X-ray emission.

More recently the outburst of XTE J1118+108 was modelled in terms of a diffusion model instead of the DIM (Wood et al. 2001). Rather than assuming that an outburst is caused by a build-up of mass in the disc during quiescence which then becomes unstable (as in the DIM), this diffusion model is based on the assumption that sporadic/variable mass transfer from the companion takes place. While a FRED morphology for the first of the two X-ray peaks of XTE J1118+108 was generated by this model, as well as the non-FRED second peak, it should also be noted that this model does not yet include a trigger and as such is currently incomplete.

The most desirable model (or perhaps combination of models) would not only explain all SXT and LHXT outbursts, but would also be consistent with models for the state changes of persistent X-ray binaries such as Cyg X-1 and LMC X-3 (including the `failed state changes' of Cyg X-1 during which the source remains in the low/hard state -- Brocksopp et al. 1999). These are all supposedly the result of variable mass transfer and/or accretion; could they all be produced by the same mechanism as X-ray transient outbursts? In particular, Wilms et al. (2001) showed that the long modulation of LMC X-3 was produced, not by the assumed precession of the accretion disc, but by quasi-periodic changes in the mass transfer rate. Likewise Fender et al. (1999) suggest variable mass transfer for outbursts in Cyg X-3. We note also that, in the Be + neutron star binary systems, outbursts occur at periastron of an eccentric orbit (e.g. van Paradijs \& McClintock 1995 and references therein) -- again suggestive of some outbursts being the result of variable mass transfer rather than instabilities within the disc.

It has also been suggested that the `superoutbursts' seen in the SU UMa class of cataclysmic variable stars are the result of a `normal' disc instability outburst triggering additional mass transfer from the companion, perhaps due to increased irradiation; the DIM alone cannot provide sufficient power for an outburst of such magnitude and duration (Frank, King, Raine 1992). A similar mechanism has been suggested for XTE J1118+480 (Kuulkers 2001). However this does not explain the outburst of GRO J1719$-$24 -- in this case the `superoutburst' {\em preceded} the `normal' outburst. If irradiation is not the sole cause of the increased mass transfer then an alternative mechanism may be necessary; models of an accretion disc wind instability limit cycle or a variable Roche lobe filling factor have been suggested for LMC X-3 (Wilms et al. 2001, Wu et al. 2001) and should be considered as potential mechanisms in other sources.

The suggestion that variable mass transfer from the companion might be important for X-ray transient outbursts is not new; e.g. Hameury et al. (1986) considered the possibility of unstable mass transfer caused by illumination of the stellar companion. This was later discounted in favour of the DIM (e.g. Mineshige \& Wheeler 1989, Gontikakis \& Hameury 1993), the main problem being that the mass transfer instability is insufficient to produce the high levels of variability observed. This problem may be avoided if the broad-band spectrum is considered. If, as appears to be the case in the low/hard state of (at least) XTE J1118+480 and GX 339$-$4 (Markoff et al. 2001, 2003; Corbel \& Fender 2002), a significant proportion of the X-ray luminosity is produced via the synchrotron mechanism then acceleration within the jet is likely to be its source, thus reducing the magnitude of the required mass transfer variability. While this is particularly important in the case of the low/hard state sources, the power in the jet ejections also has significant implications on determination of the mass accretion rate in soft events. 

Furthermore SXT outbursts of XTE J1859+226, XTE J1550$-$564 and many others have been observed to pass through the low/hard state prior to softening; they have {\em not} made a direct transition from quiescence to the high/soft (or very high) state. All recent black hole SXT outbursts have shown this behaviour and those less recent sources which have not appeared to exhibit this behaviour have not actually  been observed in sufficient detail to be certain either way (Brocksopp et al. 2002). There has been no attempt to fit the DIM to any source with the inclusion of this initial hard state and it is important that the power requirements of the initial hard state (and its associated jet) are incorporated into models.

With the low/hard state shown to be such a ubiquitous behaviour of black hole X-ray binaries it is extremely important that these LHXT `mini-outbursts' and the initial hard state prior to a black hole SXT outburst are not forgotten about. Does the same mechanism produce both the off--LHS and LHS--VHS transitions (i.e. LHS=low/hard state and VHS=very high state)? Can we assume that the off--LHS transition is produced by the same mechanism regardless of whether the outburst is a black hole SXT or LHXT? (We note that it is not yet certain whether or not the `off' and `low/hard' states are actually two distinct spectral states.) What conditions are present to cause a source to then make the LHS--VHS transition as well? These questions are particularly interesting for a source such as GS 1354$-$64 which has displayed both types of outburst.

In summary, while we do not intend this paper to necessarily favour one model over others, the low/hard state outbursts studied in this paper do not appear to be explained in terms of the DIM as it stands currently; either the DIM needs significant modification to include the properties of the jet and broad-band spectrum, or some new model is required, in which case a sporadic mass transfer model (such as the diffusion model of Wood et al. 2001) may be a viable alternative, subject to further testing and inclusion of the jet and outburst trigger. Furthermore, given the ubiquity of low/hard state behaviour in black hole X-ray binaries, it is possible that some form of variable mass transfer model could apply to black hole SXT outbusts as well, at least in providing the off--LHS transition at the onset of the outburst. If this is the case then we need to determine the conditions and mechanism (disc instability?) by which a black hole SXT then proceeds to the very high state.

\subsection{XTE J1550$-$564: A test case}

The points raised in the previous sections can be well-illustrated by using XTE J1550$-$564 as a case-study. It is particularly useful since its observing coverage has been excellent at all frequencies and all behaviours studied in this paper have been observed in this single source.

XTE J1550$-$564 was discovered when it entered a SXT outburst in 1998, passing through all of the known X-ray spectral states over the course of this outburst, including an initial low/hard state phase  (e.g. Homan et al. 2000). The outburst has been studied at X-ray--radio wavelengths and a mass transfer instability invoked as the trigger of the extremely hard ``spike'', which took place at the onset of the outburst (Wu et al. 2002). 

Radio observations revealed a jet with apparent superluminal motion, as well as the ubiquitous synchrotron spectrum (Hannikainen et al. 2001). More recent observations have shown that more than two years after the event the ejecta were still resolvable at radio {\em and at X-ray} wavelengths and with a broad-band synchrotron spectrum (Corbel et al. 2002).

Furthermore there were additional outbursts in 2000, 2001 and 2002; an initial low/hard state was observed in 2000 before the source softened and the two more recent outbursts remained in the low/hard state throughout (Belloni et al. 2002). Radio observations in 2002 confirmed the presence of a typical low/hard state jet spectrum (Corbel et al. 2002). The QPO behaviour during the 2000 outburst was particularly well-observed; both rise and fall of QPO frequency were seen during the LHS phases of the outburst. The frequency reached a plateau briefly during the LHS--VHS transition before disappearing for most of the VHS and reappearing again during the transition back to the LHS (Reilly et al 2001).

Thus study of this one source suggests (confirms?) that (i) there are problems in explaining the power-law X-ray emission with conventional outburst models, (ii) the proposed broad-band jet spectrum can be imaged and (iii) a single source can enter outbursts of different spectral properties.

\section{Conclusions}
We have accumulated multiwavelength datasets for the eight black hole X-ray binaries which have been observed to enter an outburst but remained in the low/hard state throughout. We show that, despite very similar X-ray spectral properties, the lightcurve morphologies of these sources are very different, as are the relationships between the X-ray, optical and radio lightcurves. However, the QPO and jet properties are comparable in each system.

This comparison of the low/hard state transients suggests that some mechanism to explain the outbursts in the context of their spectral properties is necessary; the disc instability model currently lacks the flexibility to do this and needs modification in order to allow for the the observed jet. We also suggest that in light of the more recent observations of both low/hard state and soft X-ray transients (e.g. XTE J1550$-$564), the various models of sporadic mass transfer should not be ruled out, at least during the low/hard state phase of the outburst -- testing of these suggestions is work in progress.

\section*{Acknowledgements}
We are extremely grateful to the many people who have contributed the various flux conversions and data for this paper -- particularly Guy Pooley, Erik Kuulkers, Mike McCollough and Jean in't Zand. The {\it RXTE}/ASM data used in this paper was obtained from the public ASM database; the {\it CGRO}/BATSE data from the web pages of the Compton Science Support Center. The Green Bank Interferometer was operated by the National Radio Astronomy Observatory for the U.S. Naval Observatory and the Naval Research laboratory during the time-period of these observations. Finally, this research has made much use of NASA's Astrophysics Data System Abstract Service and its {\sc dexter} software.

CB acknowledges a PPARC grant and STARLINK facilities at JMU. A portion of this work was performed while RMB was supported by a National Research Council Research Associateship at the Naval Research Laboratory.

\newpage

\section*{Captions}

Table 1: List of sources which entered an outburst but remained in the low/hard state throughout. Bullet points indicate the frequencies at which observations were attempted during the outburst.\\

Table 2: Conversion table for instrumental count rates to Crab units.\\

Fig. 1: Lightcurves for six of the low/hard state transient sources. The top left plot shows X-ray lightcurves for A1524$-$62, 4U 1543$-$475 and GRS 1737$-$31. The other three plots show X-ray, optical and radio lightcurves for V404 Cyg (top right), GRO J0422+32 (bottom left) and GRO J1719$-$24 (bottom right). Data have been obtained from the sources listed in Table~\ref{sources}. Arrows represent upper limits. Thick vertical lines on the X-ray panels indicate the epochs plotted in Fig.~\ref{spectrum}.\\

Fig. 1: (Contd.) X-ray, optical and radio lightcurves for GS 1354$-$64 (left) and XTE J1118+480 (right). Data have been obtained from the sources listed in Table~\ref{sources}. Arrows represent upper limits; note that the optical error-bars are smaller than the symbols in the case of GS 1354$-$64. Thick vertical lines on the X-ray panels indicate the epochs plotted in Fig.~\ref{spectrum}.\\

Fig. 2: The variability of the low frequency QPO in GRO J1719$-$24, GRO J0422+32, GS 1354$-$64 and XTE J1118+480. The increase in frequency does not correlate well with the X-ray lightcurve. Figures have been adapted from van der Hooft (1996, 1999), Brocksopp et al. (2001) and Wood et al. (2000).\\

Fig. 3: Broad-band spectra for V404 Cyg, GRO J1719$-$24, GRO J0422+32 and GS 1354$-$64, which qualitatively resemble Fig. 1 of Markoff et al. (2001). Three epochs are plotted corresponding, where possible (i.e. {\em not} in the case of GS 1354$-$64) to outburst peak (cross), secondary peak (star) and low luminosity (square) but the level of variability appears to have little effect on the shape of the spectrum. Optical data have been de-reddened as described in the text. An additional point has been included in the GRO J0422+32 plot, which is the 10$\mu$m detection by van Paradijs et al. (1994). Errors on the points are smaller than the size of the symbols (for those points for which error information is available) and downwards arrows indicate upper limits.\\

\vspace*{30cm}

\newpage
\begin{deluxetable}{llccccl}
\rotate
\tablenum{1}
\tablecolumns{7}
\tablewidth{18.5cm}
\tabletypesize{\footnotesize}
\tablehead{ 
\colhead{Source} & \colhead{Alternative Name} & \colhead{ Date} & \colhead{X-ray$^{\dagger}$} & \colhead{Optical} & \colhead{Radio} & \colhead{Data Sources}
}
\startdata
V404 Cyg&GS 2023+338&1989&$\bullet$&$\bullet$&$\bullet$&Terada et al. 1994, in't Zand et al. 1992, Wagner et\\ 
&&&&&&al. 1991, Casares et al. 1991, Han \& Hjellming 1992\\ 
A1524$-$62&TrA X-1, V*KY TrA&1990&$\bullet$&&&Barret et al. 1992, 1995\\
4U 1543$-$475&V*IL Lup &1992&$\bullet$&&&BATSE archive\\
GRO J0422+32&Nova Per, V518 Per&1992&$\bullet$&$\bullet$&$\bullet$&BATSE archive, Sunyaev et al. 1994,\\
&&&&&&Garcia et al. 1996, Shrader et al. 1994\\
GRO J1719$-$24&GRS 1716$-$249&1993&$\bullet$&$\bullet$&$\bullet$&BATSE archive, Sunyaev et al. 1994, Masetti et al.\\
&Nova Oph, V2293 Oph&&&&&1996, Hjellming et al. 1996, Della Valle et al. 1994\\
GRS 1737$-$31&&1997&$\bullet$&&&{\sl RXTE}/ASM archive, Trudolyubov et al. 1999\\
GS 1354$-$64&Cen X-2, BW Cir&1997&$\bullet$&$\bullet$&$\bullet$&{\sl RXTE}/ASM  \& BATSE archives, Brocksopp et al. 2001\\
XTE J1118+480&&2000&$\bullet$&$\bullet$&$\bullet$&{\sl RXTE}/ASM archive, Kuulkers 2001, Pooley 2001, \\
&&&&&&Wood et al. 2000\\
\tableline
\enddata
\tablecomments{$^{\dagger}$ Each source was observed at both low ($\le 12$ keV) and high ($\ge 12$ keV) X-ray energies with the exception of 4U 1543$-$475 ($\ge 20$ keV only) and XTE J1118+480 ($\le 10$ keV only). Observing coverage of the spectral range for each source is shown in Fig.~\ref{lightcurve1} and discussed in Section 1.}
\label{sources}
\end{deluxetable}

\begin{deluxetable}{llll}
\tablenum{2}
\tablecolumns{4}
\tablewidth{13cm}
\tabletypesize{\footnotesize}
\tablehead{ 
\colhead{Instrument} & \colhead{Energy Range (keV)} & \colhead{1 Crab} & \colhead{References}
}
\startdata
{\sl ROSAT}/PSPC&0.1--2.4&303 cts/s&Pietsch et al. 1993\\
{\sl GRANAT}/SIGMA&35--75&0.108 ph/s/cm$^{2}$&Trudolyubov et al. 1999\\
{\sl GRANAT}/SIGMA&75--150&0.044 ph/s/cm$^{2}$&Trudolyubov et al. 1999\\
{\sl Ginga}/ASM&1--20&2.7 cts/s/cm$^{2}$&Tsunemi et al. 1989\\
{\sl Kvant}/TTM&2--10 &1.44 cts/s/cm$^{2}$&J. in 't Zand, priv. comm.\\
{\sl Kvant}/TTM&10-27 &0.32 cts/s/cm$^{2}$&J. in 't Zand, priv. comm.\\
{\sl CGRO}/BATSE&20--100&0.3 ph/s/cm$^{2}$&M.McCollough, priv. comm.\\
{\sl RXTE}/ASM&2--10&75cts/s&Levine et al. 1996\\
{\sl ARGOS}/USA&2-12&3300 cts/s&P. Ray, priv. comm.\\
\tableline
\enddata
\label{conversions}
\end{deluxetable}

\newpage

\begin{figure*}[h]
\begin{center}
\leavevmode  
\includegraphics[height=24cm, angle=0]{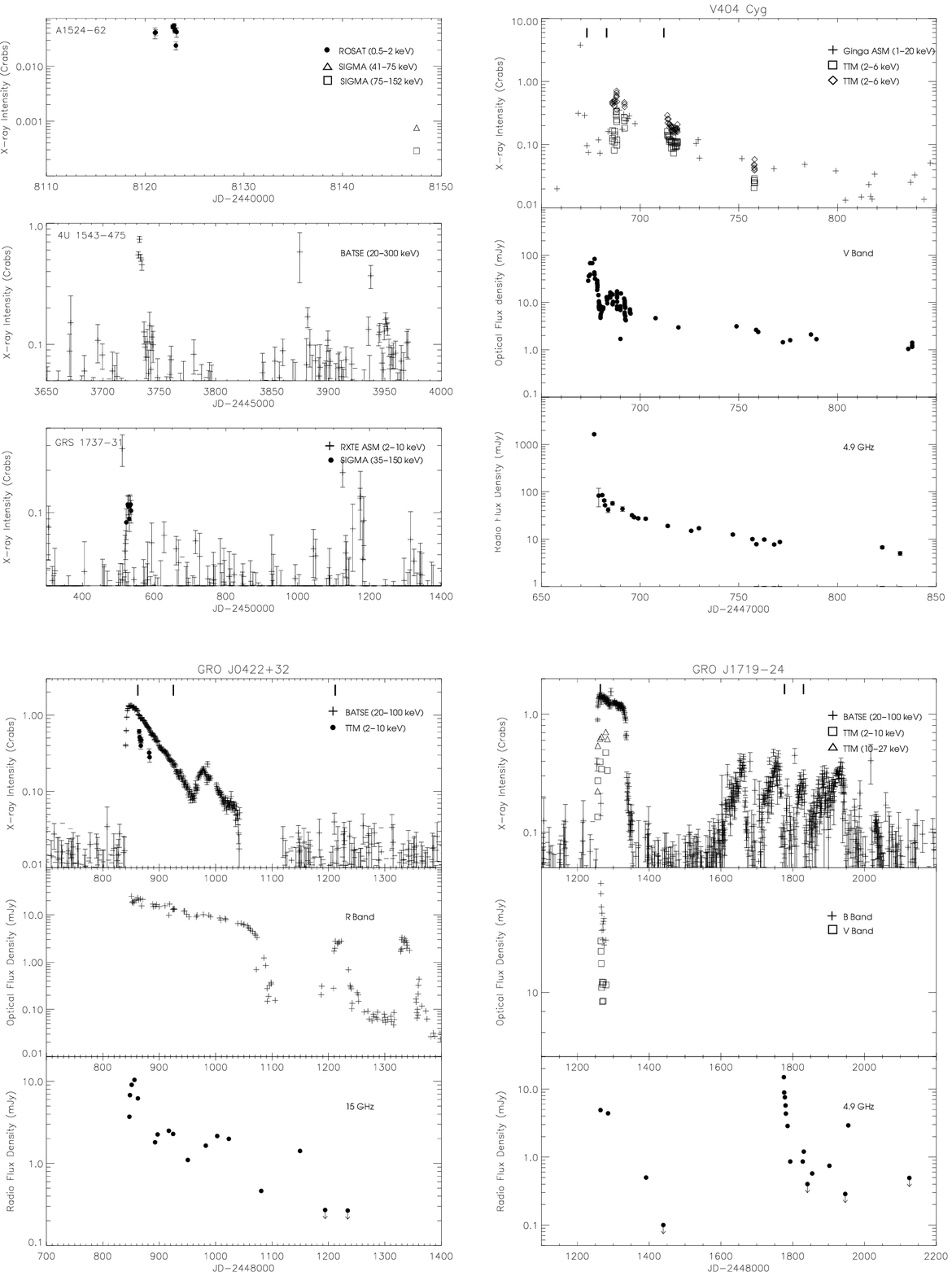} 
\caption{}
\label{lightcurve1}
\end{center}
\end{figure*}

\newpage

\begin{figure*}
\begin{center}
\leavevmode
\setcounter{figure}{0}
\includegraphics[height=9cm, angle=0]{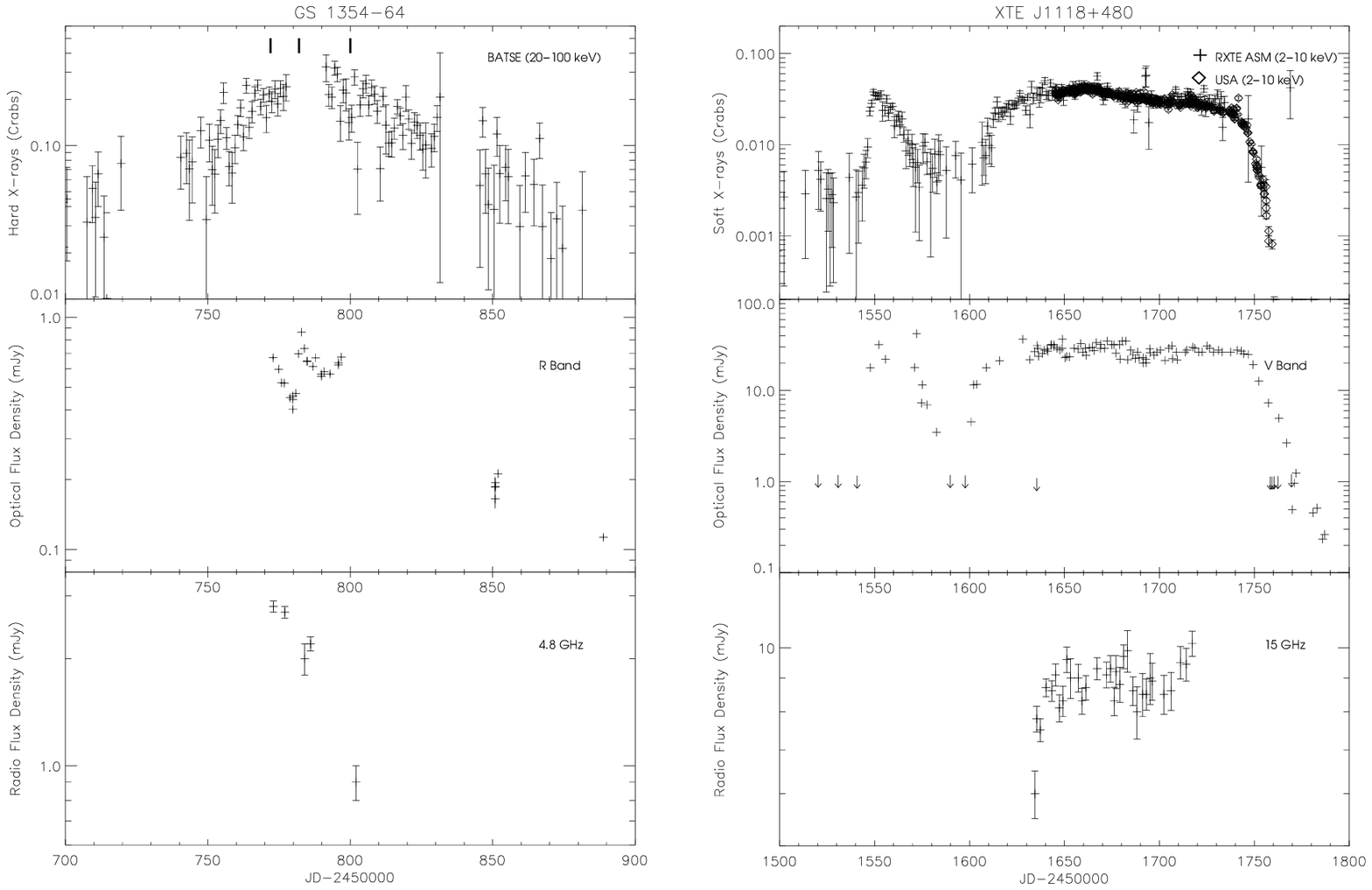} 
\caption{(Contd.)}
\label{lightcurve2}
\end{center}
\end{figure*}

\newpage

\begin{figure*}
\begin{center}
\leavevmode
\includegraphics[height=15cm, angle=90]{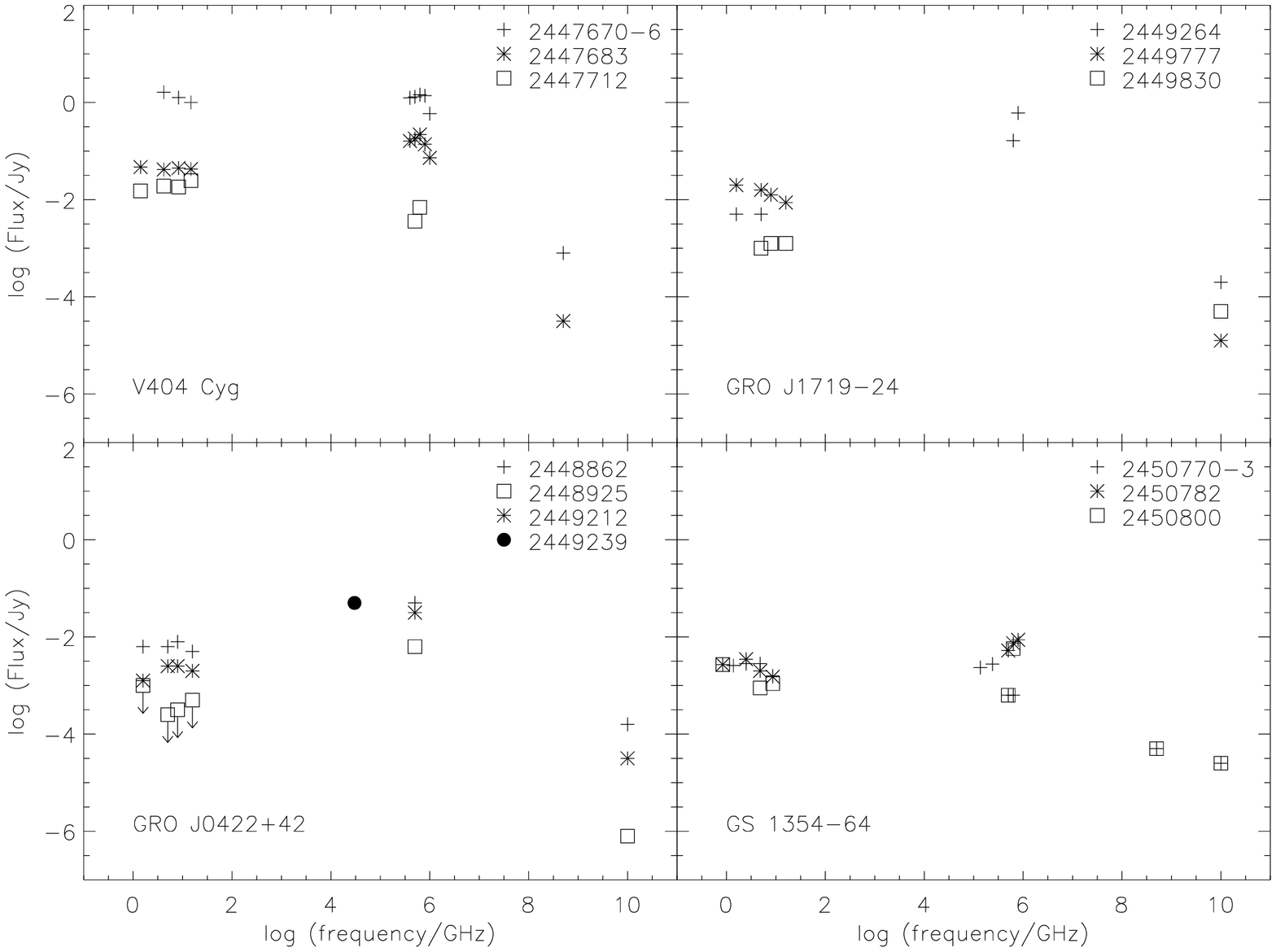}
\caption{}
\label{spectrum}
\end{center}
\end{figure*}

\newpage

\begin{figure}
\begin{minipage}{4in}
\includegraphics[width=5.5cm, angle=90]{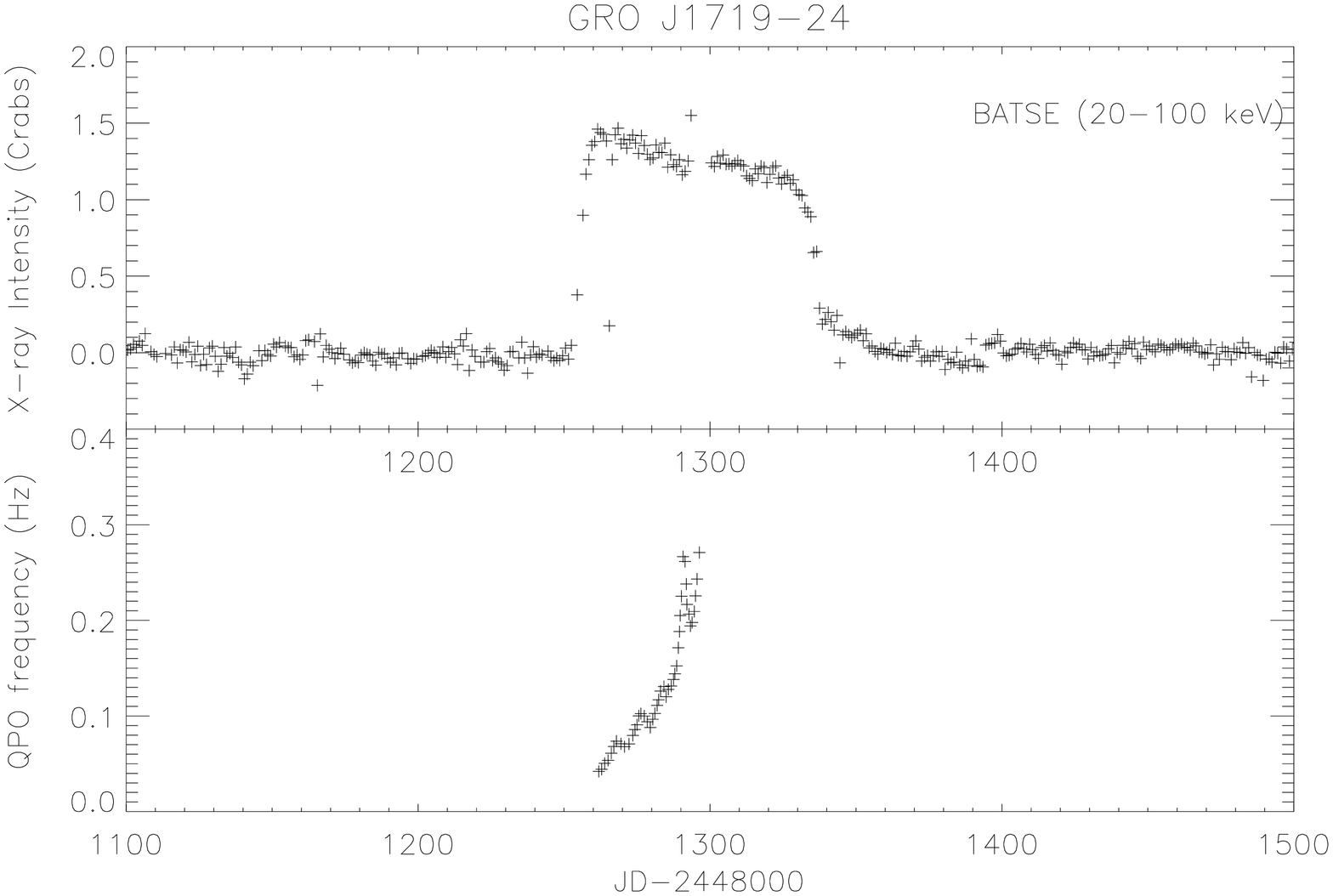}
\includegraphics[width=5.5cm, angle=90]{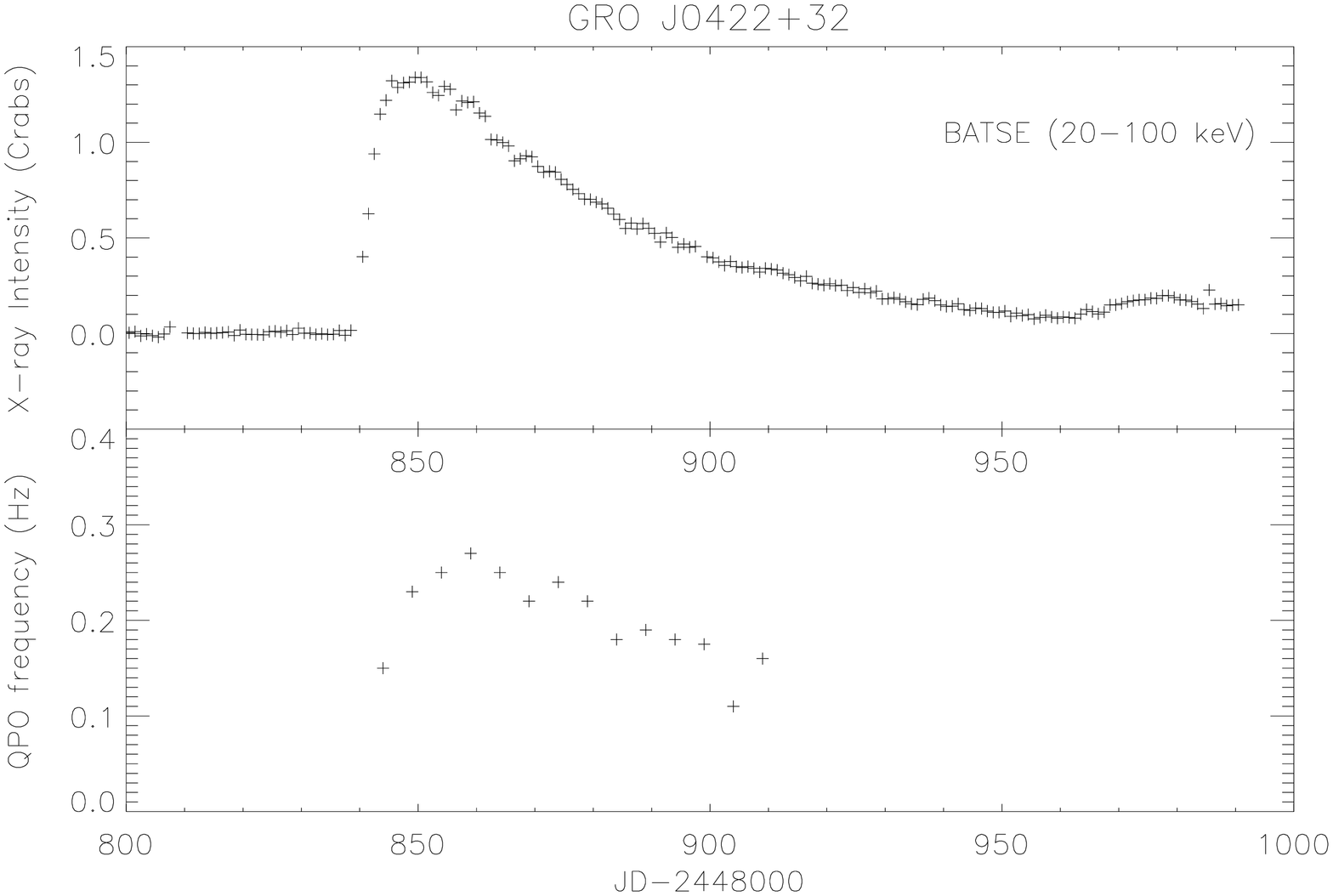}
\includegraphics[width=5.5cm, angle=90]{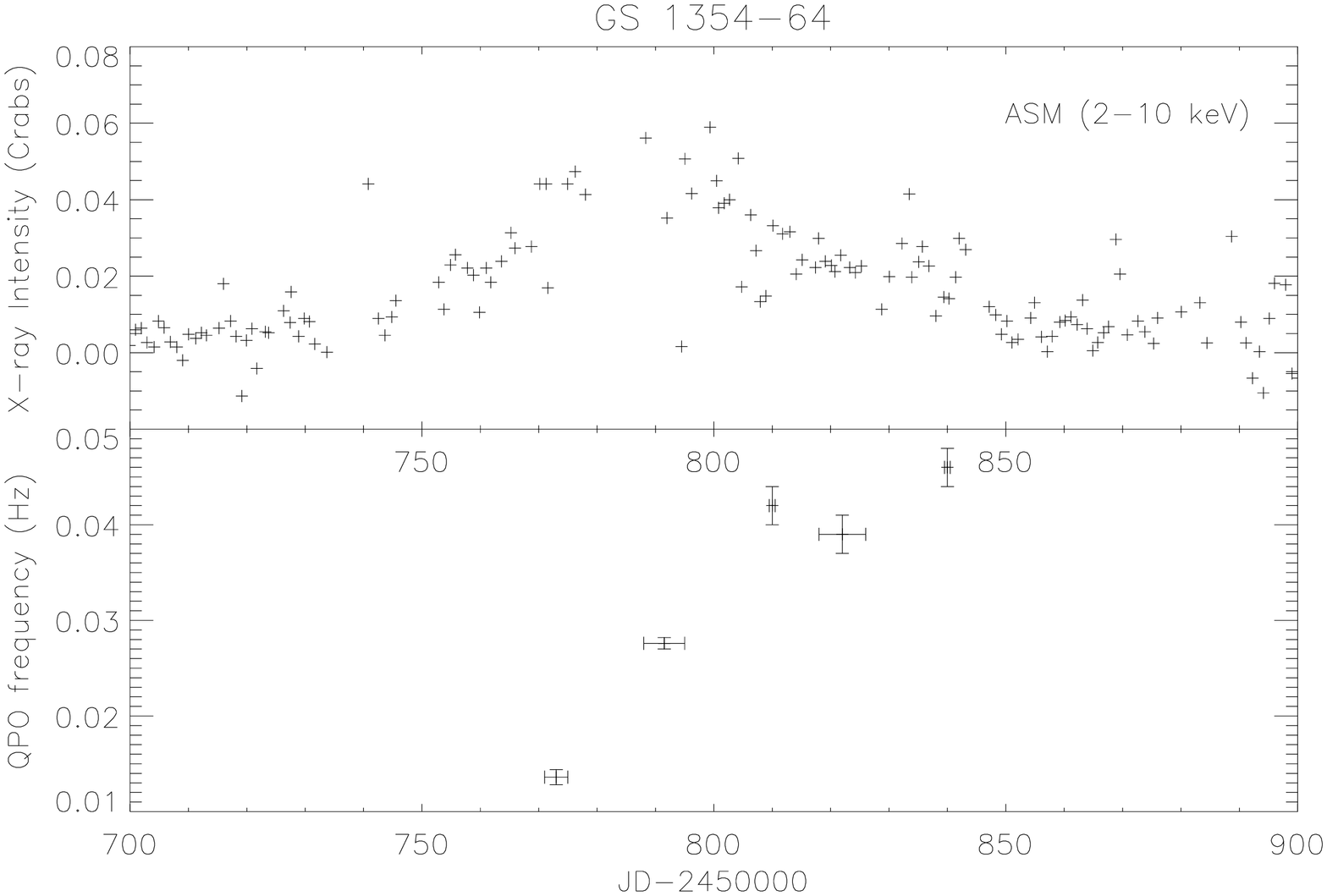}
\includegraphics[width=5.5cm, angle=90]{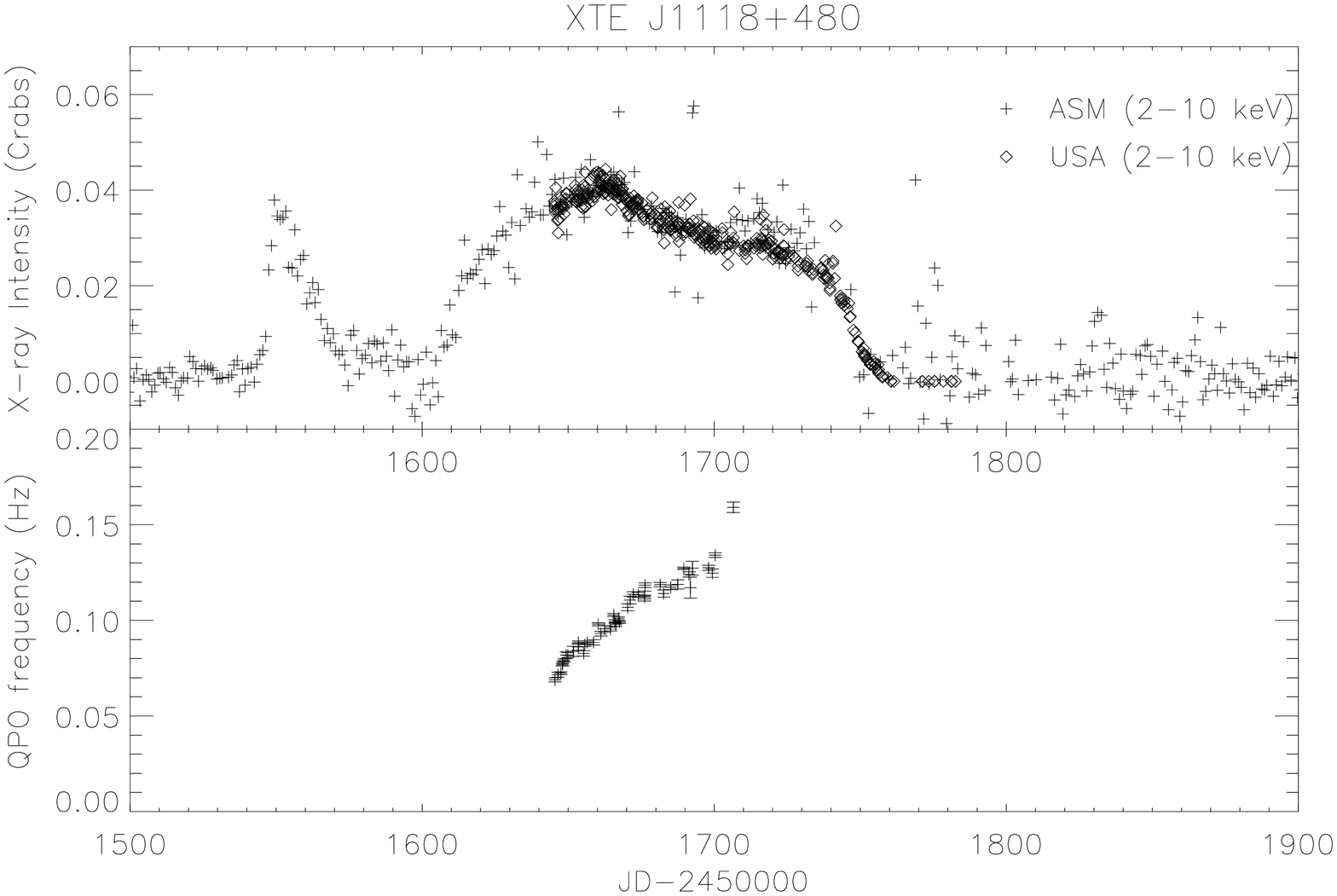}
\end{minipage} 
\label{qpo}
\caption{}
\end{figure}


\begin{references}

\reference{}Ballet J., Denis M., Gilfanov M., Sunyaev R., 1993, IAUC 5784
\reference{}Barret D. et al., 1992, ApJ, 392, L19
\reference{}Barret D., Motch C., Pietsch W., Voges W., 1995, A\&A, 296, 459
\reference{}Belloni T., Columbo A.P., Homan J., Campana S., van der Klis M., 2002, A\&A, 390, 199
\reference{}Brocksopp C., Fender R.P., Lariononv V., Lyuty V.M., Tarasov A.E., Pooley G.G., Paciesas W.S., Roche P., 1999, MNRAS, 309, 1063
\reference{}Brocksopp C., Jonker P.J., Fender R.P., Groot P.G., van der Klis M., Tingay S.J., 2001, MNRAS, 323, 517
\reference{}Brocksopp C. et al., 2002, MNRAS, 331, 765
\reference{}Callanan P.J. et al., 1995, ApJ, 441, 786
\reference{}Casares J., Charles P.A., Jones D.H.P., Rutten R.G.M, Callanan P.J., 1991, MNRAS 250, 712
\reference{}Castro-Tirado A.J., Pavlenko E.P., Shlyapnikov A.A., Brandt S., Lund N., 1993, A\&A, 276, L37
 Ortiz, J. L.
\reference{}Chen W., Shrader C.R., Livio M., 1997, ApJ, 491, 312
\reference{}Corbel S. et al., 2001, ApJ, 554, 43
\reference{}Corbel S., Fender R.P., 2002, ApJ, 573, 35
\reference{}Corbel S., Fender R.P., Tzioumis A., Tomsick J.A., Orosz J.A., Miller J.M., Wijnands R., Kaaret P., 2002, Science, 298, 196
\reference{}Corbel S., Nowak M.A., Fender R.P., Tzioumis A., Markoff S., 2003, A\&A, 400, 1007
\reference{}Cui W., Heindl W.A., Swank J.H., Smith D.M., Morgan E.H., Remillard R., Marshall F.E., 1997, ApJ, 487, L73
\reference{}Cui W., Shrader C.R., Haswell C.A., Hynes R.I., 2000, ApJ, 538, 307
\reference{}Della Valle M., Mirabel I.F., Rodr\'{\i}guez L.F., 1994, 290, 803
\reference{}Dubus G., Hameury J-M., Lasota J-P, 2001, A\&A, 373, 251
\reference{}Esin A.A., McClintock J.E., Narayan R., 1997, ApJ, 489, 865 
\reference{}Fender R.P., 2001, MNRAS 322, 31
\reference{}Fender R.P., Hanson M.M., Pooley G.G., 1999, MNRAS, 308, 473
\reference{}Fender R.P., Hendry M.A., 2000, MNRAS, 317, 1 
\reference{}Fender R.P., Hjellming R.M., Tilanus R.P.J., Pooley G.G., Deane J.R., Ogley R.N., Spencer R.E., 2001, MNRAS, 322, L23
\reference{}Frank J., King A., Raine D., 1992, `Accretion Power in Astrophysics', CUP
\reference{}Garcia M.R., Callanan P.J., McClintock J.E., Zhao P., 1996, ApJ, 460, 932
\reference{}Hameury  J.M., King A.R., Lasota J.P., 1986, 162, 71
\reference{}Han X., Hjellming R.M., 1992, ApJ, 400, 304
\reference{}Hannikainen D. et al., 2001, ESA SP-459, p291, Proceedings of the Fourth INTEGRAL Workshop, 4-8 September 2000, Alicante, Spain. Eds. A. Gimenez, V. Reglero, C. Winkler. 
\reference{}Harmon B.A., Paciesas W.S., Fishman G.J., 1993, IAUC 5874
\reference{}Harmon B.A., Wilson R.B., Finger M.H., Paciesas W.S., Rubin B.C., Fishman G.J., 1992, IAUC 5504
\reference{}Harmon B.A. et al., 1995, Nature, 374, 703
\reference{}Heise J., 1997, IAUC 6606
\reference{}Hjellming R.M., Rupen M.P., Shrader C.R., Campbell-Wilson D., Hunstead R.W., McKay D.J., 1996, ApJ, 470, L105
\reference{}Homan J., Wijnands R., van der Klis M., Belloni T., van Paradijs J., Klein-Woldt M., Fender R., Mendez M., 2000, ApJS, 132, 377
\reference{}Hynes R.I., Haswell C.A., 1999, MNRAS, 303, 101
\reference{}Hynes R.I., Mauche C.W., Haswell C.A., Shrader C.R., Cui W., Chaty S., 2000, ApJ, 539, L37
\reference{}Hynes R.I., Zurita C., Haswell C.A., Casares J. Charles P.A., Pavlenko E.P., Shugarov S.Y., Lott D.A., 2002, MNRAS, 330, 1009
\reference{}In't Zand J.J.M., Pan H.C., Bleeker J.A.M., Skinner G.K., Gilfanov M.R., Sunyaev R.A., 1992, A\&A, 266, 283
\reference{}Kaluzienski L.J., Holt S.S., Boldt E.A., Serlemitsos, Eadie G., Pounds K.A., Ricketts M.J., Watson M., 1975, ApJ, 201, L121
\reference{}Kanbach G., Straubmeier C., Spruit H.C., Belloni T., 2001, Nature, 414, 180
\reference{}King A.R., Ritter H., 1998, MNRAS, 293, L42
\reference{}Kitamoto S., Miyamoto S., Tsunemi H., Makishima M., 1984, PASJ, 36, 799
\reference{}Kuulkers E., 2001, AN, 322, 9
\reference{}Lasota J-P., 2001, New Astron. Rev., 45, 449 
\reference{}Levine et al., 1996, ApJ, 469, L33
\reference{}Liu Q.Z., van Paradijs J., van den Heuvel E.P.J., 2001, A\&A, 368, 1021
\reference{}Markoff S., Falcke H., Fender R.P., 2001, A\&A, 372, L25
\reference{}Markoff S., Nowak M.A., Corbel S., Fender R.P., Falcke H., 2003, A\&A, 397, 645
\reference{}Masetti N., Bianchini A., Bonibaker J., Della Valle M., Vio R., 1996, A\&A, 314, 123
\reference{}Matilsky T.A., Giacconi R., Gursky H., Kellogg E.M., Tanabaum H.D., 1972, ApJ, 174, L53
\reference{}McClintock J.E. et al., 2001, ApJ, 555, 477
\reference{}Miller J.M., Remillard R.A., 2002, IAUC 7920
\reference{}Mineshige S., Wheeler J.C., 1989, ApJ, 343, 241
\reference{}Oosterbroek T. et al., 1997, A\&A, 321, 776
\reference{}Paciesas W.S., Briggs M.S., Harmon B.A., Wilson R B., Finger M.H., 1992, IAUC 5580
\reference{}Pietsch W., Haberl F., Gehrels N., Petre R., 1993, A\&A, 273, L11
\reference{}Pooley G.G., 2001, MNRAS, 324, L23
\reference{}Reilly K.T. et al., 2001, ApJ, 561, L183
\reference{}Remillard R., Morgan E., Smith D., Smith E., 2000, IAUC 7389
\reference{}Revnivtsev M.G., Borozdin K.N., Priedhorsky W.C., Vikhlinin A., 2000b, ApJ, 530, 955
\reference{}Revnivtsev M.G., Sunyaev R., Borozdin K.N., 2000a, A\&A, 361, L37
\reference{}Revnivtsev M.G. et al., 1998, A\&A, 331, 557
\reference{}Shabaz T., Ringwald F.A., Bunn J.C., Naylor T., Charles P.A., Casares J., 1994, MNRAS, 271, L10
\reference{}Shrader C.R., Wagner R.M., Hjellming R.M., Han X.H., Starrfield S.G., 1994, 434, 698
\reference{}Soria R., Wu K., Hannikainen D., McCollough M., Hunstead R., 2001, Proceedings of the JHU/LHEA Workshop: "X-ray Emission from Accretion onto Black Holes" (Baltimore, June 20--23, 2001), (astro-ph/0108084)
\reference{}Spruit H.C., Kanbach G., 2002, A\&A, 391, 225
\reference{}Sunyaev R.A. et al., 1994, AstL, 20, 777
\reference{}Tanaka Y, Lewin W.H.G., 1995, `X-Ray Binaries' Chap 3, CUP, Eds. Lewin W.H.G., Van Paradijs J., Van den Heuvel E.
\reference{}Terada K., Miyamoto S., Kitamoto S., Egoshi W., PASJ, 46, 677
\reference{}Trudolyubov S. et al., 1999, A\&A, 342, 496
\reference{}Tsunemi H., Kitamoto S., Manabe M., Miyamoto S., Yamashita K., Nakagawa M., 1989, PASJ, 41, 391
\reference{}Ueda Y. et al., 1997, IAUC 6627 
\reference{}Uemura M. et al., 2000, PASJ, 52, L15
\reference{}Van der Hooft F., et al. 1996, ApJ, 458, L75
\reference{} Van der Hooft F., et al. 1999, ApJ, 513, 477
\reference{} Van der Klis M., 1995, `X-ray Binaries', Chapter 6, Eds. Lewin W.H.G., van Paradijs J., van den Heuvel E., CUP
\reference{} Van Paradijs J., McClintock J.E., 1995, `X-Ray Binaries' Chap 1, CUP, Eds. Lewin W.H.G., Van Paradijs J., Van den Heuvel E.
\reference{} Van Paradijs J., Telesco C.M., Kouveliotou C., Fishman G.J., 1994, 429, L19
\reference{}Wagner R.M., Starrfield S.G., Howell S.B., Kreidl T.J., Bus S.J., Cassatella A., Bertram R., Fried R., 1991, ApJ, 378, 293
\reference{}Whittet D.C.B., 1992, `Dust in the Galactic Environment', IOP: New York
\reference{}Wijnands R., van der Klis M., 1999, 514, 939
\reference{}Wilms J., Nowak M.A., Pottschmidt K., Heindl W.A., Dove J.B., Begelman M.C., 2001, MNRAS, 320, 327 
\reference{}Wood K.S. et al., 2000, ApJ, 544, L45
\reference{}Wood K.S., Titarchuk L., Ray P.S., Wolff M.T., Lovellette M.N., Bandyopadhyay R., 2001, ApJ, 563, 246
\reference{}Wu K., Soria R., Page M.J., Sakelliou I., Kahn S.M., de Vries C.P., 2001, A\&A, 2001, 365, L267
\reference{}Wu K., Soria R., Campbell-Wilson D., Hannikainen D., Harmon B.A., Hunstead R., Johnston H., McCollough M., McIntyre V., 2002, ApJ, 565, 1161
\reference{}Zhang S.N., Cui W., Harmon B.A., Paciesas W.S., Remillard R.E., van Paradijs J., 1997, ApJ, 477, L95 
\reference{}\.Zycki P.T., Done C., Smith D.A., 1999, MNRAS, 309, 561

\end{references}
\end{document}